\documentclass[twocolumn]{aastex7}

\usepackage{xcolor}
\usepackage{colortbl}
\usepackage{amsmath}

\begin{document}

\title{The exozodi spectral effect: Residual habitable zone dust may bias exoEarth characterization} 

\correspondingauthor{Miles Currie}
\email{mcurrie.astro@gmail.com}

\author[0000-0003-3429-4142]{Miles H. Currie}
\email{mcurrie.astro@gmail.com}
\affiliation{NASA Goddard Space Flight Center, Greenbelt, MD 20771, USA}
\author[]{Christopher C. Stark}
\email{christopher.c.stark@nasa.gov}
\affiliation{NASA Goddard Space Flight Center, Greenbelt, MD 20771, USA}
\author[0000-0002-0006-1175]{Eleonora Alei}
\email{eleonora.alei@nasa.gov}
\affiliation{NASA Goddard Space Flight Center, Greenbelt, MD 20771, USA}
\author[0000-0002-2989-3725]{Aki Roberge}
\email{aki.roberge-1@nasa.gov}
\affiliation{NASA Goddard Space Flight Center, Greenbelt, MD 20771, USA}

\begin{abstract}
All exoplanetary systems are expected to host exozodiacal dust, or exozodi, originating from planetesimals. For many stars, exozodi will likely be the largest source of astrophysical noise in direct observations of terrestrial exoplanets. Nearby Sun-like systems likely have $\gtrsim3$x more habitable zone (HZ) dust than our solar system, which must be removed from direct images and spectra to reveal exoplanet signals. The albedo of this micron-sized dust varies smoothly over VIS---NIR wavelengths, but exozodi can be composed of different materials that can impact its color. If left unsubtracted, exozodiacal light will add cloud-like continuum emission to extracted spectra, potentially biasing characterization studies by reducing the apparent absorption depth of spectral features. To quantify these effects, we simulate exoEarth systems with a range of exozodi densities and compositions, and apply an atmospheric retrieval tool to synthetic Habitable Worlds Observatory (HWO) spectra. We find that exozodi at levels similar to the solar system (i.e., 1 zodi) can reduce the apparent depth of visible wavelength molecular absorption features by up to 50\%, an effect that worsens at longer wavelengths. To measure molecular abundances, significant post-processing may be required to remove exozodi to a fractional residual that tightens with dust density. However, targeting a binary detection result for an absorbing species instead relaxes this requirement by an order of magnitude, especially at higher spectral resolution. Understanding and mitigating the effects of exozodi in extracted exoEarth spectra is critical to characterize HZ exoplanet environments with HWO and ultimately to search for signs of habitability and life.
\end{abstract}


\section{Introduction}\label{sec:intro}

A primary goal of the Habitable Worlds Observatory (HWO), a NASA flagship mission concept recommended for development by the Astro2020 Decadal Survey \citep{committeeforadecadalsurveyonastronomyandastrophysics2020astro2020PathwaysDiscoveryAstronomy2023}, is to find and study $\sim 25$ potentially Earth-like planets in nearby FGK systems. This will involve probing the planets' atmospheres for signs of habitability and life in high-contrast direct spectra spanning the NUV \citep[e.g.,][]{ahmed2025ultraviolet} to the NIR \citep[][]{krissansen2025wavelength}. Precursor science and trade-space studies for HWO are now underway \citep[e.g.,][]{belikov2024coronagraph, feinberg2024habitable, stark2024paths, tuchow2024hpic}. An important part of this preparatory work is to understand the possible systematic and astrophysical limitations for the observatory. Although we have control over the observatory architecture design and technology development, the astrophysical scene is intrinsic to a given target system. Emission from circumstellar dust in exoplanet systems will be blended with planet light and will likely be the largest source of astrophysical noise in HWO observations of Earth-like exoplanets \citep[][]{robergeExozodiacalDustProblem2012}. We must therefore develop techniques to account for its effect on observations. 

Dust in the habitable zone (HZ) of an exoplanet system is analogous to zodiacal light---the solar system dust that resides within a few AU of the Sun---and as such is referred to as exozodiacal dust, or ``exozodi''\citep[e.g.,][]{kralExozodiacalCloudsHot2017, wyatt2025theory}. Zodiacal dust in the solar system is composed of $\sim1-100$~micron-sized carbonaceous/silicate grains \citep[][]{Grun1985}, spanning a wide range of temperatures, from $\sim$150~K near the asteroid belt to several hundred~K and hotter toward the solar corona. Exozodi, on the other hand, is broadly defined as an analogous dust population in other stellar systems \citep[][]{kralExozodiacalCloudsHot2017, wyatt2025theory}; for this work we are particularly interested in warm (T~$\sim300$~K) dust, and use the term ``exozodi'' to refer to this HZ dust component. 
The composition of exozodi in a typical system is not generally known, but the presence of exozodi has been detected in nearby Sun-like systems via thermal excess measurements \citep[e.g.,][]{beichmanNewDebrisDisks2006, kennedy2013bright, patel2014sensitive, mennesson2014constraining, ertelHOSTSSurveyExozodiacal2020}. 

While there is still much to be learned about exozodi, modeling and observations continue to improve and provide some constraints on its properties \citep[e.g.,][]{kralExozodiacalCloudsHot2017, wyatt2025theory}. Observationally, the HOSTS collaboration has surveyed nearby Sun-like stars for exozodi, and found that the median system contains $\sim3$ zodis of dust, where one ``zodi'' is equivalent to the vertical optical depth of the solar system HZ dust \citep{ertelHOSTSSurveyExozodiacal2020}. Exozodi may also be spatially inhomogeneous, potentially exhibiting clumps, gaps, or warps in its morphology \citep{starkDetectabilityExoEarthsSuperEarths2008, starkNEWALGORITHMSELFCONSISTENT2009}. This is expected considering that inhomogeneities are observed when dust gravitationally interacts with planets in our own solar system \citep[][]{jackson1989solar,dermottCircumsolarRingAsteroidal1994, reachObservationalConfirmationCircumsolar1995}. Since a large fraction of nearby Sun-like stars are expected to be dusty \citep[][]{beichmanNewDebrisDisks2006, ertelHOSTSSurveyExozodiacal2020}, exozodi is a concern for direct observations of terrestrial HZ exoplanets, even at dust densities comparable to the zodiacal dust in our solar system \citep[][]{defrereNullingInterferometryImpact2010, defrereDirectImagingExoEarths2012, robergeExozodiacalDustProblem2012}. 

Thus, emission from exozodiacal dust will likely need to be removed from future direct observations of exoEarths, which may be challenging depending on the density, orientation, and whether any asymmetries are present \citep[e.g.,][]{kammererSimulatingReflectedLight2022, currieMitigatingWorstcaseExozodiacal2023, mennesson2024current}. 
ExoEarth yield models \citep[e.g.,][]{morgan2023exo, stark2024paths} and direct imaging exposure time calculators \citep[e.g.,][]{robinsonCharacterizingRockyGaseous2016, lustig-yaegerCoronagraphTelescopeNoise2019} currently assume exozodi can be subtracted to the Poisson noise limit. For spatially smooth exozodi, a simple approach is to fit toy models (radial/azimuthal polynomials) to subtract the exozodi \citep[][]{kammererSimulatingReflectedLight2022}. This method was demonstrated to subtract up to 1000 zodis of warm dust to the Poisson noise limit in simulated images of face-on systems, or up to 50 zodis for systems $60^\circ$ inclined from face-on \citep[][]{kammererSimulatingReflectedLight2022}. 

However, if exozodi is spatially inhomogeneous due to significant mean motion resonance structure, removal becomes challenging beyond a few zodis worth of dust \citep[][]{currieMitigatingWorstcaseExozodiacal2023}. Our ability to remove exozodi is strongly tied to both the density of the dust and the inclination of the system due to the forward scattering of starlight \citep[e.g.,][]{stark2014revealing, stark2015pseudo}. Recent simulations of exozodi removal techniques perform poorly for highly inclined or edge-on systems, and have only been tested at $0.5~\mu$m \citep[][]{kammererSimulatingReflectedLight2022, currieMitigatingWorstcaseExozodiacal2023}. 

To search for signs of habitability and life in nearby FGK systems, HWO observations will target key atmospheric gases, including H$_2$O, O$_2$, O$_3$, CH$_4$, and CO$_2$, all of which have one or more absorption bands in the nominal $\sim$0.2--1.7 $\mu$m wavelength range of HWO. Measuring the abundances of these gases is necessary to interpret them within the planetary environment and understand whether biosignatures are present \citep[][]{schwietermanOverviewExoplanetBiosignatures2024}. 
State-of-the-art atmospheric retrieval frameworks provide a promising method for detecting gases and measuring their abundances in exoEarth atmospheres \citep[e.g.,][]{damianoReflectedSpectroscopySmall2022, latouf2023bayesian, gilbert-janizekRetrievedAtmospheresInferred2024, youngInferringChemicalDisequilibrium2024}. However, modern retrieval algorithms also assume exozodi emission can be completely removed from observed direct spectra and only contributes additional Poisson noise. It is currently unclear how residual exozodi emission in extracted exoEarth spectra may bias atmospheric retrievals, or whether these algorithms must be updated to consider exozodi within the suite of retrieved parameters. 

In the NUV--VIS--NIR wavelength range, exozodi is approximately a gray scatterer of stellar light \citep{leinert1997ReferenceDiffuse1998}. To first order, the two effects on the extracted planetary spectrum are: (1) increased noise for all spectral bins \citep[][]{robergeExozodiacalDustProblem2012} and (2) brighter continuum if the exozodi is not fully subtracted, causing a cloud-like effect on the observed spectrum. Beyond first order effects, exozodi may have chromaticity like its colder debris disk counterparts \citep[e.g.][]{debesColorGradientsDetected2008, renDebrisDiskColor2023}, and its impact on an exoplanet spectrum may therefore be wavelength dependent. 

In this work, we investigate how exozodi may affect our ability to characterize the atmospheres of Earth-like exoplanets. We present our models and methods in Section~\ref{sec:theproblem}, quantify the exozodi spectral effect on extracted exoplanet spectra and atmospheric retrievals in Section~\ref{sec:quantify}, contextualize our findings in Section~\ref{sec:mitigation}, and conclude in Section~\ref{sec:conclusion}. 

\section{Methods}\label{sec:theproblem}

\subsection{Exozodi models}\label{sec:exozodimodels}

\begin{figure}
    \centering
    \includegraphics[width=\linewidth]{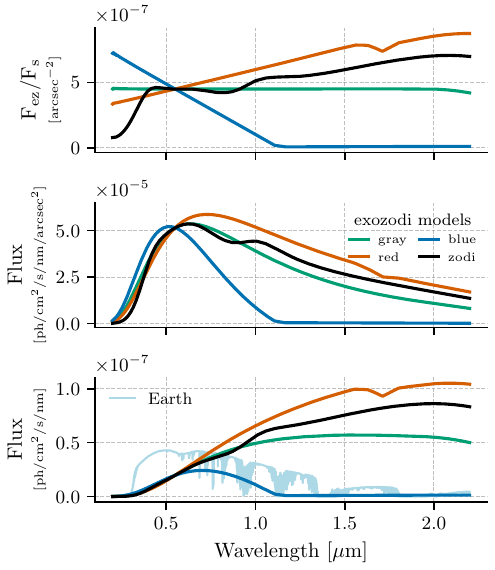}
    \caption{Models of exozodiacal dust exhibiting gray, red, blue, and solar system zodi-like colors, each shown for 3 zodis of dust viewed at $60^\circ$ from face-on. The top and middle rows are the exozodi contrast and flux per area, respectively. The bottom row shows the total wavelength-dependent exozodi flux captured within a photometric aperture of size 0.7 $\lambda$/D for a 7.2~m HWO aperture. The thin light-blue line in the bottom panel shows the flux of a modern Earth-twin planet ($F_\mathrm{p}$) at 10~pc for direct comparison; the vertical separation between an exozodi curve and the Earth curve in a given spectral bin is the exozodi-to-planet flux ratio $F_\mathrm{ez}/F_\mathrm{p}$ at that wavelength. The total exozodi flux integrated over a photometric aperture for a given spectral bin depends on the color of the exozodi.}
    \label{fig:ez_models}
\end{figure}

The physical properties of exozodiacal dust are not well characterized, and it may not have the same composition, grain size distribution, or grain shape as zodiacal dust. We therefore adopt a range of models to simulate the effect of exozodi on future direct imaging observations. As a first order approximation, we consider simple cases such as a true gray scatterer and the empirical Solar System zodiacal dust profile \citep[e.g. \S 8 of][]{leinert1997ReferenceDiffuse1998}. We additionally consider plausible exozodi models that exhibit color by using debris disk observations as analogs for dust composition in the HZ (plotted in Figure~\ref{fig:ez_models}). Most debris disks tend to be blue in color; we use the empirical color relations derived by \citet{renDebrisDiskColor2023} to construct a model for blue exozodi. Blue scattering is typically caused by small particles, reflecting the ongoing collisional cascades occurring in debris disks \citep[e.g.][]{renDebrisDiskColor2023}. Small grains can be blown out from the system by radiation pressure, but the rate is highly dependent on dust composition and porosity. Our blue exozodi model contains grains with plausible size/composition properties for dust that does not rapidly blow out \citep{arnoldEffectDustComposition2019a, lebretonIcyKuiperBelt2012}. While less common, debris disks can also appear red in color; to model red exozodi we use the observed color of the debris disk HD~15115 \citep{debesColorGradientsDetected2008}.

Regardless of scattering properties, exozodiacal dust, like solar system dust, is an extended source and will add extra flux to the background. The total exozodi flux integrated over a fixed photometric aperture is invariant to system distance: the flux from individual dust grains in the disk decreases as $d^{-2}$, but the number of dust grains in the photometric core increases as $d^2$. 

While the intrinsic surface brightness of exozodi depends on its color and density, the \emph{observed} exozodi flux also depends on the size of the telescope's point spread function (PSF), which is proportional to $\lambda/D$. Thus, the exozodi flux within the PSF, blended with the planet signal, will increase with wavelength even for a simple gray scatterer. We illustrate this in the lower panel of Figure~\ref{fig:ez_models}, where we plot the total observed exozodi flux for a fixed photometric aperture of size 0.7 $\lambda/D$ and assume a 7.2-m HWO aperture. 


To explore the possibility that nearby FGK systems may have more dust than our solar system \citep[][]{ertelHOSTSSurveyExozodiacal2020}, we define a variable N$_\mathrm{ez}$ that parameterizes the density of exozodi relative to our solar system's zodiacal dust. For this work, we define one ``zodi'' of exozodi as equal to the brightness of the solar system's zodiacal dust in the V-band as seen face-on. Of course, this brightness can be spatially inhomogeneous for any single system due to forward scattering, even if the disk-integrated exozodi density is fixed. 
For a fixed exozodi density, edge-on orientations will appear brightest due to forward scattering. 



\subsection{Planetary models}\label{sec:planetmodels}

For simplicity and efficiency, we use the radiative transfer capability of the \texttt{rfast} framework \citep[][]{robinsonExploringValidatingExoplanet2023} to simulate clear and cloudy (50\% cloud fraction) modern Earth-like spectral models. Because we are testing the specific effect of exozodi, and not the fidelity of a climate/photochemical model, the post-industrial Earth ICRCCM 62 atmosphere model packaged with \texttt{rfast} is sufficient for our purposes. We use the built-in radiative transfer process to calculate model spectra, but omit the noise calculation process in \texttt{rfast} and substitute a more sophisticated noise model developed for HWO (see Section~\ref{sec:etc}). 

The modern Earth atmosphere we consider has trace amounts of CO$_2$ and CH$_4$ (400 ppm and 1 ppm, respectively), which may be challenging to detect with HWO even for cases with no exozodi. Therefore, we also consider an Archean Earth atmosphere \citep[1\% CO$_2$ and 0.2\% CH$_4$;][]{arneyPaleOrangeDots2017, arneyOrganicHazeBiosignature2018} for comparison. In contrast, the Archean atmosphere has only a trace abundance of O$_2$, compared with the 21\% O$_2$ of the modern Earth atmosphere.

\subsection{Data simulation and noise model}\label{sec:retrievals}

\subsubsection{Exposure Time Calculator}\label{sec:etc}
To synthesize realistic HWO spectral data for atmospheric retrievals, we use the spectroscopy mode of the \texttt{pyEDITH} exposure time calculator \citep[ETC;][]{pyedith2025, alei2026multibandpass}. \texttt{pyEDITH} is a Python-based ETC with heritage from the Altruistic Yield Optimizer \citep[AYO;][]{starkMAXIMIZINGExoEarthCANDIDATE2014, stark2024paths}, commissioned by the HWO Technology Maturation Project Office as the standard, publicly available coronagraphic ETC for observation planning. The code interfaces with HWO engineering specifications\footnote{\url{https://github.com/HWO-GOMAP-Working-Groups/Sci-Eng-Interface/}} to calculate exposure times that are consistent with the telescope and coronagraph capabilities of the HWO exploratory mission architectures.  


To effectively explore the wavelength trade space, we assume six spectral channels between  0.4 and 1.7 $\mu$m. 
Unless otherwise noted, we adopt $R = 140$ for $\lambda_\mathrm{VIS} < 1\mu$m and $R=40$ for $\lambda_\mathrm{NIR} > 1\mu$m. To maintain consistency with the definition of one ``zodi'' of dust in our synthetic exoplanetary systems, we assume each one has a Sun-like star with one solar radius and $T_\mathrm{eff} = 5780$~K. The exoplanets are assumed to be at 1 AU from the host star, and are one Earth radius in size, with modern or Archean Earth-like atmospheres (Section~\ref{sec:planetmodels}). We focus on target systems located 10 pc from Earth. In addition to the \texttt{pyEDITH} parameters mentioned, we adopt the default HWO architecture values included in \texttt{pyEDITH} for all other observatory-related parameters \citep[][]{pyedith2025}.

For this work, to account for imperfect exozodi subtraction, we refactor the exozodi-related count rates and noise floors from the pyEDITH standard equation of \citet{pyedith2025}. This new feature is available in the ETC for public use.

\subsubsection{Signal-to-noise model}

We consider a simple S/N equation adopted from \citet[][]{starkExoEarthYieldLandscape2019}, where S/N is related to the total exposure time and count rates of the planet, exozodi, and other background sources:
\begin{equation}\label{eq:SNR_simple}
    \mathrm{S/N} = \sqrt{ t_\mathrm{exp} \frac{C_\mathrm{p}^2 - \mathrm{(S/N)}^2C_\mathrm{nf}^2}{C_\mathrm{p} + C_\mathrm{b} + C_\mathrm{ez}}},
\end{equation}
where S/N is the signal-to-noise ratio of a given spectral bin, $t_\mathrm{exp}$ is the total science exposure time, and $C_\mathrm{p}$, $C_\mathrm{nf}$, $C_\mathrm{ez}$, $C_\mathrm{b}$  are the photon count rates within a photometric aperture on the detector of the planet, systematic noise floor, exozodi, and all other background sources (stellar, detector, thermal, etc.), respectively.  

Ignoring the noise floor ($C_{nf} = 0$) and assuming the exozodi can be perfectly subtracted to the Poisson noise limit, exposure time requirements will increase relative to the exozodi photon count. Using Equation~\ref{eq:SNR_simple}, the exposure time accounting for exozodi is
\begin{equation}
    t_\mathrm{exp, ez} = t_\mathrm{exp, 0}\left( 1 + \frac{C_\mathrm{ez}}{C_\mathrm{p} + C_\mathrm{b}} \right),
\end{equation}
where $t_\mathrm{exp, ez}$ is the total exposure time adjusted for additional exozodi Poisson noise and $t_\mathrm{exp, 0}$ is the exposure time of the system without an exozodi contribution (i.e., $C_\mathrm{ez} = 0$). 
Since exozodi is to first order a gray scatterer, the faintest portions of the planetary reflectance spectrum (e.g., the deepest parts of molecular absorption bands) will be most affected by this additional noise. 

Background flux that cannot be subtracted to the Poisson noise limit will introduce a systematic noise floor such that $\mathrm{S/N} > 1$ cannot be achieved for any source dimmer than the noise floor ($C_\mathrm{p} < C_\mathrm{nf}$). This concept has been well explored for stellar leakage, where coronagraphic speckles cannot be perfectly subtracted for any given exposure \citep[e.g.][]{redmond2024exoplanet}. It is challenging to estimate the stellar leakage noise floor ($C_\mathrm{nf,s}$) without extensive lab experiments for a particular optical system, so it is useful to introduce the concept of a post-processing factor ($\mathrm{PPF}_s$) that scales down the count rate of the stellar leakage $C_s$ to estimate the stellar leakage noise floor, given by
\begin{equation}
    C_\mathrm{nf,s} = C_s / \mathrm{PPF}_s.
\end{equation}

If exozodi cannot be fully subtracted to the Poisson noise limit, due to, for example, spatial variations \citep{currieMitigatingWorstcaseExozodiacal2023}, this introduces an exozodi noise floor $C_\mathrm{nf,ez}$ in addition to the stellar leakage noise floor. Following convention, we introduce an \textit{exozodi} postprocessing factor $\mathrm{PPF}_\mathrm{ez}$ such that 
\begin{equation}
    C_\mathrm{nf,ez} = C_\mathrm{ez} / \mathrm{PPF}_\mathrm{ez}.
\end{equation}
To tie these concepts back to Equation~\ref{eq:SNR_simple}, we extend the single-noise-floor framework of \citet{starkExoEarthYieldLandscape2019} to include a separate exozodi noise-floor term: residual exozodi after imperfect subtraction acts as a systematic signal bias on the planet measurement, analogous to residual stellar leakage. Because both terms represent deterministic residual biases rather than stochastic noise, we combine them linearly:
\begin{equation}
    C_\mathrm{nf} = C_\mathrm{nf,s} + C_\mathrm{nf,ez}.
\end{equation}

The exozodi post-processing factor $\mathrm{PPF}_\mathrm{ez}$ accounts for imperfect exozodi subtraction and directly affects the noise floor. Practically, this ``leftover'' exozodi can contaminate flux measurements of the target. This manifests when some amount of exozodi flux is not removed during the fitting process due, for example, to an unknown ``clump'' of dust or a model mismatch near the planet. Critically, there may also be cases where there is residual exozodi flux within the planet's PSF even if the exozodi appears to have been fully subtracted nearby after fitting. 

\begin{figure*}
    \centering
    \includegraphics[width=\linewidth]{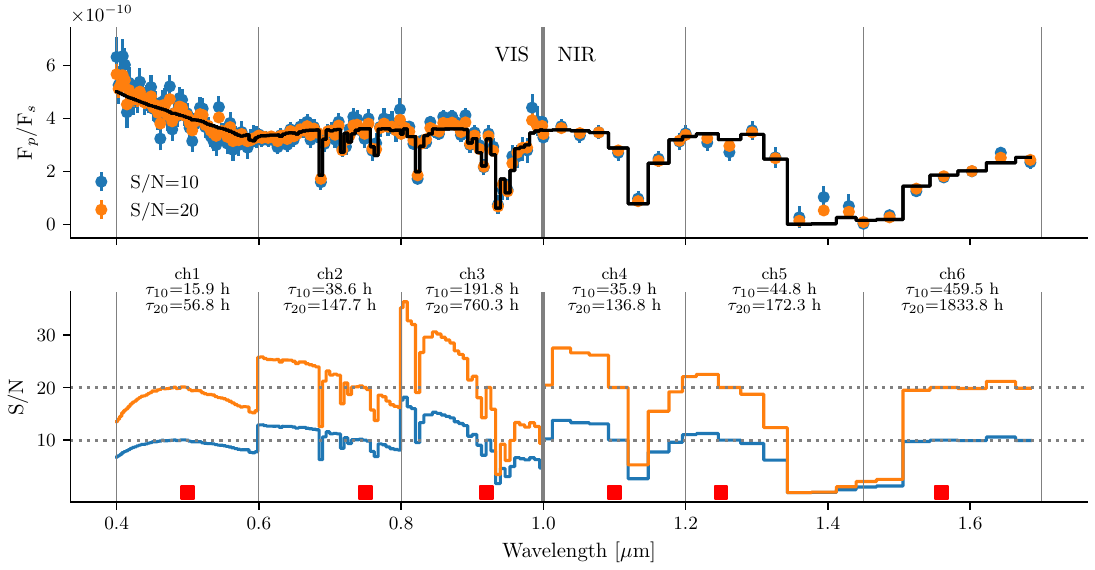}
    \caption{An example of synthetic HWO Earth spectra at 10 pc generated with \texttt{pyEDITH} (upper) and the corresponding S/N values for each wavelength bin (lower). The gray vertical lines separate the spectral channels assumed for this work. Exposure times were fixed such that S/N=10 or S/N=20 is achieved at reference wavelength bins (red squares) for each spectral channel. We placed the reference bins at continuum locations away from the deepest molecular absorption features in each channel, so that exposure times are set by the continuum S/N requirement rather than by the depth at a specific absorption line.}
    \label{fig:snr}
\end{figure*}

To simulate data for spectral retrieval analysis, we use \texttt{pyEDITH} to calculate signal-to-noise (S/N) values for each spectral bin. Figure~\ref{fig:snr} shows example synthetic spectra of Earth at 10~pc (upper panel), and the corresponding S/N for each wavelength bin (lower panel), calculated using Equation~\ref{eq:SNR_simple}. We assign a ``reference wavelength bin'' in each spectral channel (red square markers in Figure~\ref{fig:snr}), and fix the exposure time for each channel as the value needed to achieve S/N=10 or 20 in the reference bins. Practically, this results in different exposure times for each spectral channel, also listed in Figure~\ref{fig:snr}. 

\subsubsection{Retrieval Model}\label{sec:retrievalmodel}
To more realistically quantify how exozodi may impact our ability to characterize an exoEarth atmosphere, we simulate observations of Earth-like exoplanets and run a retrieval algorithm to estimate planetary characteristics. To test how a current atmospheric retrieval model responds to dusty exoplanet systems, we apply the default \texttt{rfast} algorithm \citep[][]{robinsonExploringValidatingExoplanet2023} to our synthesized data. Using \texttt{emcee} \citep[][]{foreman2013emcee}, the \texttt{rfast} algorithm explores the parameter space of exoplanet properties for $10^6$ likelihood evaluations, generating forward models to compare with the input data and building Bayesian posterior distributions. We include a comprehensive set of exoplanet parameters, with the following priors:
\begin{equation}\label{eq:params}
    \mathcal{P}(\mathbf{\omega})=
\begin{cases}
\log_{10}(f_{O_2}~[\mathrm{VMR}]) \sim \mathcal{U}(-10, 0) \\
\log_{10}(f_{H_2O}~[\mathrm{VMR}])  \sim \mathcal{U}(-10, 0) \\
\log_{10}(f_{CO_2}~[\mathrm{VMR}])  \sim \mathcal{U}(-10, 0) \\
\log_{10}(f_{CH_4}~[\mathrm{VMR}])  \sim \mathcal{U}(-10, 0) \\
T_0~[\mathrm{K}] \sim \mathcal{U}(273, 647) \\
\log_{10}(P_0~[\mathrm{bar}])  \sim \mathcal{U}(0, 8)\\
\log_{10}(A_s)  \sim \mathcal{U}(-2, 0) \\
\log_{10}(R_p~[\mathrm{R_\oplus}])  \sim \mathcal{U}(-1, 1) \\
\log_{10}(g_p~[\mathrm{m~s^{-2}}]) \sim \mathcal{U}(0, \log_{10}25)  \\
\log_{10}(\Delta P_\mathrm{cld}~[\mathrm{bar}])  \sim \mathcal{U}(0, 8) \\
\log_{10}(P_\mathrm{t,cld}~[\mathrm{bar}]) \sim \mathcal{U}(0, 8) \\
\log_{10}(\tau_\mathrm{cld}) \sim \mathcal{U}(-3,3)  \\ 
\log_{10}(f_\mathrm{cld})  \sim \mathcal{U}(-3,0)  \\
\end{cases}
\end{equation}
where $\mathbf{\omega}$ is the forward model parameter state vector, $\mathcal{U}(\mathrm{lower}, \mathrm{upper})$ represents a uniform distribution with finite probability between the lower and upper bounds; the units for each parameter appear in square brackets. $f_{mol}$ is the volume mixing ratio (VMR) of a molecule. $T_0$, $P_0$, and $A_s$ are the surface temperature, pressure, and albedo of the planet. $R_p$ and $g_p$ are the radius and surface gravity of the planet. Finally, clouds are parameterized using the thickness ($\Delta P_\mathrm{cld}$), cloud-top pressure ($P_\mathrm{t,cld}$), cloud optical depth ($\tau_\mathrm{cld}$), and fraction of planetary surface obscured ($f_\mathrm{cld}$).

\section{Results}\label{sec:quantify}
In this section, we quantify the exozodi spectral effect and discuss the implications for observations, molecular detection, and characterization.

\subsection{Impact on continuum}

\begin{figure*}
    \centering
    \includegraphics[width=\linewidth]{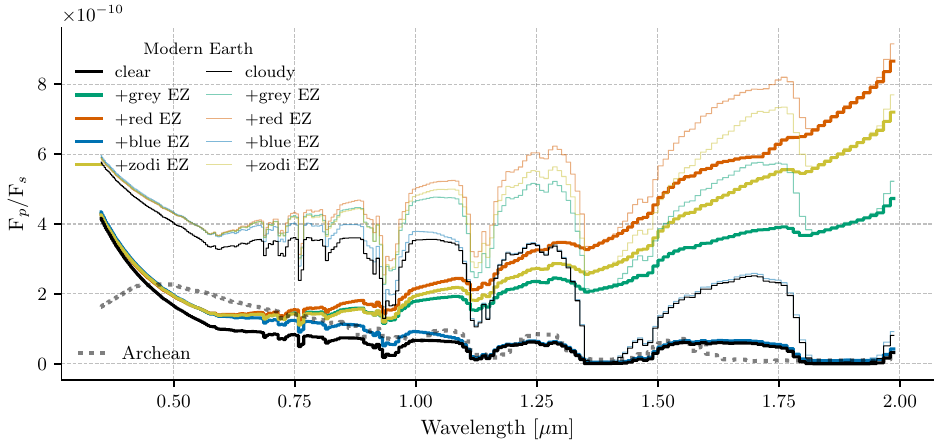}
    \caption{Modern Earth spectra for clear (thick lines) and cloudy (f$_\mathrm{cld} = 0.5$, thin lines) atmospheres, and the effect of adding 1 zodi of exozodiacal dust (colors) with an assumed 0.7 $\lambda$/D photometric aperture. Adding exozodi to these spectra raises the apparent spectral continuum, an effect similarly achieved by adding clouds to the atmosphere.}
    \label{fig:theproblem_colors}
\end{figure*}

\begin{figure*}
    \centering
    \includegraphics[width=\linewidth]{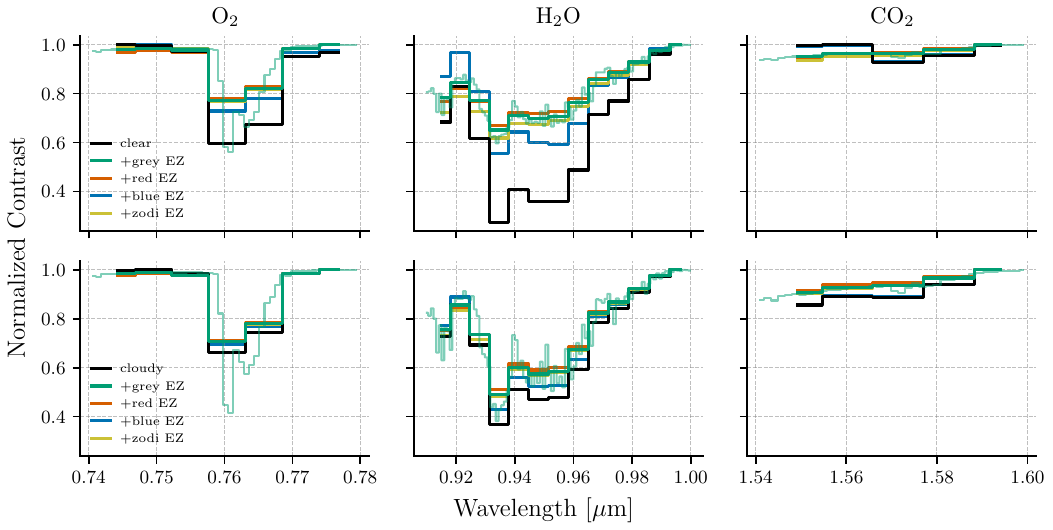}
    \caption{Normalized molecular absorption bands for a modern Earth-like exoplanet. The upper and lower rows show clear and cloudy (f$_\mathrm{cld}$) sky atmospheres, respectively. The black lines represent spectra where exozodi was perfectly removed (i.e. $\mathrm{PPF}_\mathrm{ez}$ = $\infty$). The colored lines represent the effect of adding 1 zodi of unsubtracted (i.e. $\mathrm{PPF}_\mathrm{ez}$ = 1) dust for different exozodi color cases. These spectra are nominally shown at R=140, but for comparison we include an R=1000 spectrum for the gray exozodi case as the thin line (see Section~\ref{sec:specres_disc} for a discussion on the topic of spectral resolution). Residual exozodi in the data will reduce absorption band depths, and this effect is a function of the exozodi color, how well it is subtracted from the data, and whether the exoplanet atmosphere is clear or cloudy.}
    \label{fig:theproblem_molecules}
\end{figure*}

This residual exozodi can cause the observed continuum to be elevated relative to the true planetary reflectance continuum, which can only reduce the depth of absorption bands relative to the continuum, never physically deepen them; the size of this reduction scales with the exozodi flux scattered into the aperture at a given wavelength, and can be small (e.g.\ for blue exozodi in the infrared). Figure~\ref{fig:theproblem_colors} shows Earth spectra with one zodi of dust emission for a variety of exozodi cases, where one zodi is equivalent to the vertical optical depth of the solar system HZ dust \citep{ertelHOSTSSurveyExozodiacal2020}. Clouds can also raise the continuum because they are reflective \citep[e.g.][]{kelkar2025earth}, but, unlike exozodi, clouds do not always reduce relative band depth. For optically thick bands (e.g.\ H$_2$O at 1.4 and 1.9~$\mu$m in Figure~\ref{fig:theproblem_colors}), clouds can make absorption features appear deeper rather than shallower, with the sign and magnitude of the effect depending on the band optical thickness and the underlying P--T profile. We plot selected normalized molecular absorption bands for these cases in  Figure~\ref{fig:theproblem_molecules}, considering perfect exozodi subtraction (black curve, PPF$_\mathrm{ez}$ = $\infty$) and no exozodi subtraction (colored curves, PPF$_\mathrm{ez}$ = 1). 

Additionally, exozodi may have color (Section~\ref{sec:exozodimodels}), and these noise and continuum effects will be wavelength dependent. In general, the exozodi effect is worse at longer wavelengths, and thus molecules that absorb at longer wavelengths will be more affected than those with short wavelength absorption. Furthermore, molecules with multiple absorption bands within the HWO wavelength range, like O$_2$, H$_2$O, or CH$_4$, will be particularly affected by this since the noise will be worse and the absorption band will be shallower than expected at longer wavelengths. This is a potential point of confusion for spectral retrieval algorithms that do not account for this effect. 

As expected, characteristically red exozodi has the worst effect on all plotted absorption bands (Figure~\ref{fig:theproblem_molecules}), but only by less than a few percent compared with gray and zodiacal-like dust. Characteristically blue exozodi, which does not scatter as much stellar light at wavelengths beyond $\sim0.6~\mu$m, has less of an effect on shorter wavelength absorption bands, and almost no effect on longer wavelength bands. Systems with blue dust may therefore be more observationally favorable for exoplanet atmospheric characterization than for systems with gray, red, or zodi-like dust.

The requirements on the achieved exozodi post-processing factor, PPF$_\mathrm{ez}$, will depend on the brightness of both the exozodi and the planetary target. It is therefore convenient to define a post-processed exozodi-to-planet flux ratio $(F_\mathrm{ez}^\prime/F_\mathrm{p})_V$ in the V-band, where $F_\mathrm{ez}^\prime$ is the exozodi flux in the V-band \textit{after} applying PPF$_\mathrm{ez}$. The flux of the planet, $F_\mathrm{p}$ depends on its intrinsic characteristics, orbit, and distance, and we assume it is not affected by exozodi post-processing. For conciseness, we refer to the exozodi-to-planet flux ratio $(F_\mathrm{ez}^\prime/F_\mathrm{p})_V$ as the residual exozodi contamination (REC) ratio in the text.  
The post-processed and pre-processed exozodi-to-planet flux ratios are directly related by $F_\mathrm{ez}^\prime/F_\mathrm{p} = (F_\mathrm{ez}/F_\mathrm{p})\,/\,\mathrm{PPF}_\mathrm{ez}$, where $F_\mathrm{ez}/F_\mathrm{p}$ is the ratio prior to any exozodi post-processing. This pre-processing ratio can be read directly from the bottom panel of Figure~\ref{fig:ez_models} as the vertical separation between a given exozodi curve and the Earth-twin curve, and it scales linearly with N$_\mathrm{ez}$.



\subsection{Impact on spectral features}\label{sec:abs_band_depth}

\begin{figure*}
    \centering
    \includegraphics[width=\linewidth]{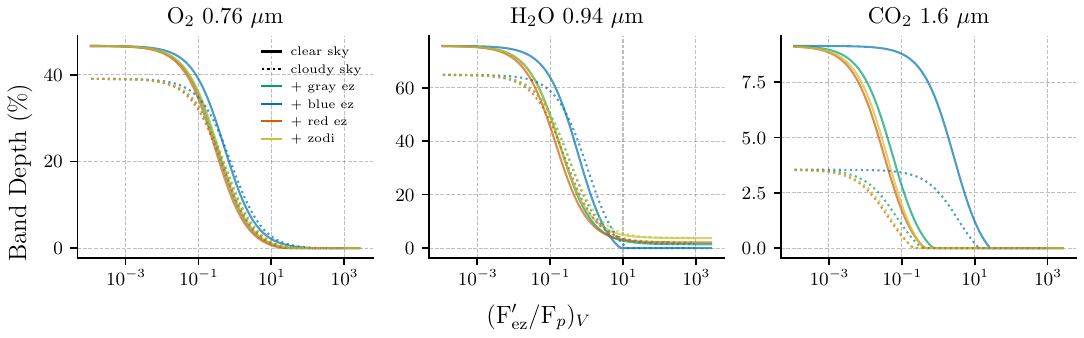}
    \caption{Molecular absorption depths as a function of residual-exozodi to planet flux ratio. For absorption bands $< 1 \mu$m, exozodi must be subtracted from the observations to F$_\mathrm{ez}^\prime$/F$_p$ $\lesssim 10^{-2}$ to preserve the full band depth. The exozodi effect is worse at longer wavelengths, and this requirement becomes stricter by nearly two orders of magnitude for the CO$_2$ band for all cases but blue exozodi. Cases with gray, red, or zodi-like dust affect the band depths to within a few percent of one another.  }  
    \label{fig:abs_depth}
\end{figure*}

We present measured molecular absorption band depths for the molecules O$_2$, H$_2$O, and CO$_2$ for clear/cloudy sky modern Earth atmospheres as a function of REC ratio in Figure~\ref{fig:abs_depth}, where different curves represent exozodi color and clear/cloudy sky exoplanet atmospheres. The maximum absorption depth asymptote can be interpreted as the exozodi having little to no effect on the spectral continuum. There is a distinct difference in the maximum absorption depths between clear and cloudy sky atmospheres, simply due to the cloud effect on absorption depths. For each molecule, the absorption depth then decreases and asymptotes to zero with increasing REC ratio, corresponding to a scenario where the exozodi is bright enough to dominate the continuum, washing out any absorption features. The curves between these asymptotes, however, encode how each of these molecular absorption bands responds to moderate amounts of exozodi. 

Each curve in Figure~\ref{fig:abs_depth} departs from the maximum absorption depth asymptote at different REC ratios, setting requirements on exozodi removal processes that may be wavelength dependent. If we only consider gray exozodi for simplicity, longer wavelength absorption bands depart the band maximum asymptote at lower REC ratios. In other words, we will require stricter exozodi subtraction routines to remove the higher levels of exozodi flux at longer wavelengths (see Section~\ref{sec:measuringabundances}). 

The zodi-like model we consider has a nearly identical effect on O$_2$ and H$_2$O as the gray exozodi, and a slightly worse effect than gray exozodi for CO$_2$. This is due to its generally gray albedo and slight reddening. If exozodiacal dust is similar to the dust in our solar system, we should expect that molecules absorbing at longer wavelengths ($>1\mu$m) will be more challenging to detect than the simple gray albedo case. 

The blue exozodi model has a milder effect on molecular absorption depths when compared to the red exozodi model. 
For molecular detectability purposes, blue exozodi represents the observationally favorable case for exoEarth characterization. Red exozodi, however, represents the opposite scenario: The higher intrinsic flux of the red exozodi is compounded by the necessary increase in photometric aperture with wavelength, and the total observed exozodi flux exceeds all other models (Figure~\ref{fig:ez_models}).

Clouds can decrease the maximum absorption depth of a molecular band relative to its continuum (e.g. Figure~\ref{fig:theproblem_molecules}), and adding exozodi to an observed spectrum can create degenerate absorption band depths for cloudy/clear sky atmospheres. This effect depends on how well the exozodi is subtracted from the observation, and how precisely the band depth can be measured. For O$_2$ and H$_2$O the highest potential for degeneracy is near the intersection of the clear and cloudy sky curves in Figure~\ref{fig:abs_depth}, where the REC ratio is greater than $0.1$. We therefore recommend that to gain cursory information about the cloudiness of an exoplanet atmosphere in the wake of exozodi, the exozodi must be subtracted from the observation such that the exozodi flux is less than one tenth the planet's flux in the V-band. This is likely achievable for targets $\sim10$ pc away with current exozodi removal methods \citep[][]{kammererSimulatingReflectedLight2022, currieMitigatingWorstcaseExozodiacal2023}. 


Unsubtracted exozodi also decreases the steepness of the Rayleigh slope by up to 33\% in the clear-sky case (26\% for cloudy) for 1 zodi of dust relative to the exozodi-free slope. Considering only the atmospheric composition, the steepness of the Rayleigh slope is dependent on the size distribution of the particles (molecules, aerosols, etc.) in the atmosphere.  The reduced apparent Rayleigh slope steepness associated with unsubtracted exozodi therefore may indicate a higher mean molecular weight than the truth, potentially biasing atmospheric retrievals.

\subsection{Exozodi post-processing requirements}

Since absorption band depth is a function of the achieved residual exozodi-to-planet flux ratio (REC ratio), we can directly translate this into requirements on exozodi post-processing. The level of post-processing required is dependent on the fidelity of information desired; for simplicity, we discriminate two classes of planetary atmospheric analyses: (1) measuring the abundance of a molecule in an atmosphere for precise environmental characterization and (2) merely detecting the presence of a molecular or spectral feature as a binary result. The former class of analysis will likely require significant post-processing to avoid biasing atmospheric retrieval algorithms, while the latter may have increased flexibility in the level of post-processing since Bayesian retrieval is not necessarily needed for simple detection algorithms. 

\subsubsection{Measuring Molecular Abundances}\label{sec:measuringabundances}
\begin{figure*}
    \centering
    \includegraphics[width=\linewidth]{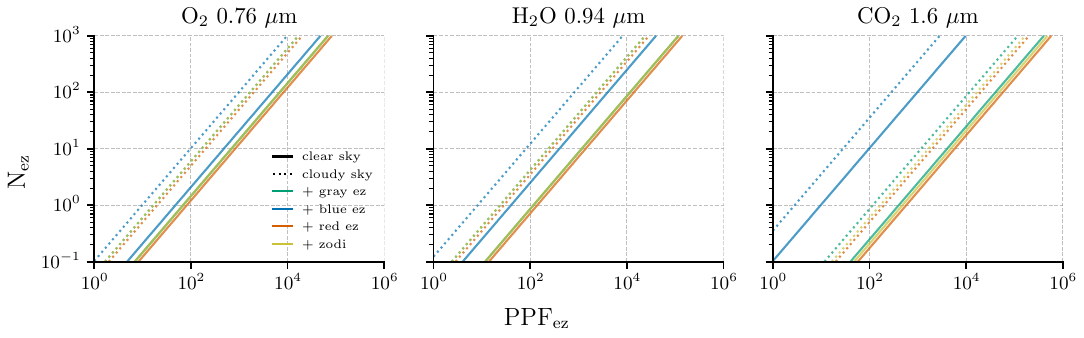}
    \caption{Contours in N$_\mathrm{zodi}$--PPF$_\mathrm{ez}$ space showing the maximum F$_\mathrm{ez}^\prime$/F$_p$ required to preserve the absorption band depth to within 1\% of its uncontaminated value in the processed data. For all but the blue exozodi case, systems with 3 zodis of dust will require post-processing factors of $>100$ for molecules $< 1\mu$m, and $>1000$ for longer wavelength absorption bands. For systems with more dust, achieving the required PPF$_\mathrm{ez}$ will become increasingly challenging.}
    \label{fig:NezPPF}
\end{figure*}

We assume that to avoid biasing retrievals, the added exozodi contamination must negligibly alter the extracted spectral continuum away from the true planetary continuum (see Section~\ref{sec:retrievalresults} for an analysis of this assumption).
For the absorption bands in Figure~\ref{fig:abs_depth}, we quantify the maximum plausible REC ratio that preserves the absorption band to within 1\% of its maximum depth as a first approximation\footnote{See Section~\ref{sec:retrievalresults} for a validation of this approximation.}, and plot the contour of this value in N$_\mathrm{ez}$--PPF$_\mathrm{ez}$ space for each case in Figure~\ref{fig:NezPPF}.  More exozodi in a given system will require stronger post-processing factors to preserve the full absorption depth of the molecular band. This requirement is slightly eased if the planetary atmosphere is cloudy, because the planet will have higher reflectivity, or for cases with bluer exozodi. However, we may in general need to achieve a post-processing factor on the order of $10^2-10^3$ for a system with just a few zodis of dust, and we discuss this potential challenge in Section~\ref{sec:mitigation}. 

Furthermore, the REC ratio is also a function of planetary brightness and thus distance.
Obviously, systems that are closer to Earth may ease exozodi post-processing requirements. For example, the PPF$_\mathrm{ez}$ requirement to achieve a given REC ratio would be 4x larger for a planet at 20 pc compared with a planet at 10 pc.  

\subsubsection{Detecting Spectral Features}\label{sec:detectingfeatures}
On the other hand, if we are interested in simply detecting the presence of a molecular absorption feature (a binary result), the post-processing requirements for exozodi removal can be relaxed. At its core, the goal for this type of analysis is to simply detect a departure from the continuum (i.e. an absorption feature) with some confidence. Thus, our ability to detect an absorption feature will depend on its apparent depth and the S/N of the spectral bins. 

Since exozodi introduces contamination that decreases absorption band depth with increasing REC ratio (Figure~\ref{fig:abs_depth}), there exists some tolerance for exozodi contamination that may be unique to a given observed band. Practically, this means that the tolerable REC ratio for simply detecting a feature may be less strict than if we require the spectral continuum to be effectively uncontaminated (i.e. for molecular abundance measurement).

At the nominal R=140 resolving power for the HWO VIS channel, detecting an absorption feature would mean that the absorption band depth must be $>40\%$ with respect to the continuum to detect the feature at 3$\sigma$ significance, if the spectral continuum bins are observed at SNR=10. Given this, O$_2$ and H$_2$O may be detectable if REC ratios of $\lesssim10^{-2}$ and $\lesssim10^{-1}$ could be achieved, respectively, but CO$_2$ would be challenging (Figure~\ref{fig:abs_depth}). In other words, a detection of O$_2$ would likely require similarly strict post-processing as that needed to measure its abundance, but targeting a simple H$_2$O detection may ease post-processing requirements by over an order of magnitude. This can be further improved by achieving a higher spectral resolution, and we discuss this in Section~\ref{sec:specres_disc}.

\subsubsection{Retrieval Results}\label{sec:retrievalresults}
\begin{figure}
    \centering
    \includegraphics[width=\linewidth]{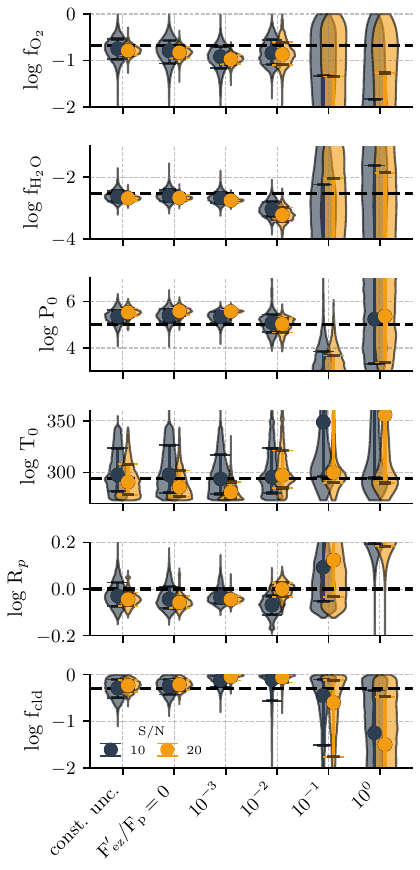}
    \caption{Selected retrieved planetary parameters as a function of F$_\mathrm{ez}^\prime$/F$_p$. With increasing F$_\mathrm{ez}^\prime$/F$_p$, retrieved molecular abundances are lower than the truth, and all parameters become unconstrained due to the increased residual exozodi flux in the data.}
    \label{fig:retrieved_params}
\end{figure}

We create a sample set of five total cases of plausible exoEarth/exozodi systems. For comparison to previous work, we include a simplistic case where all spectral bins have uncertainties fixed at the 1 $\mu$m bin's uncertainty, as in \citet{robinsonExploringValidatingExoplanet2023}, denoted as ``const. unc.'' Then, to simulate the impact of exozodi on retrievals, we calculate synthetic data for F$_\mathrm{ez}^\prime$/F$_p$ = $0$--$1$ at regular logarithmic intervals. This range of flux ratios encompasses cases where the exozodi is perfectly removed from the data (Poisson noise only) through the case where the exozodi residuals are as bright as the planet (F$_\mathrm{ez}^\prime$/F$_p$ = 1). We plot the retrieved distributions for selected parameters in Figure~\ref{fig:retrieved_params}, where each panel is a retrieved parameter, the dashed black line is the true value for the parameter, and the retrieved distributions are plotted in blue and orange (S/N = 10 and 20, respectively, at the reference wavelengths in each channel), with 16\%, 50\%, and 84\% sample percentiles as the black markers and whiskers. For reference, we also include the corner plots for all cases in Appendix~\ref{apx:cornerplots}.

Our retrieval results validate the hypothesis previously put forth in Section~\ref{sec:abs_band_depth}: at F$_\mathrm{ez}^\prime$/F$_p$ $> 10^{-3}$, it becomes more challenging to retrieve atmospheric parameters from our synthetic data compared with the zero exozodi case. We note that this is the same REC ratio where exozodi begins to effectively contaminate the absorption bands (Figure~\ref{fig:abs_depth}). In particular, the absorption bands of molecular features become shallower with increasing REC ratio, and the abundances retrieved for O$_2$ and H$_2$O are reduced until they are no longer constrained at REC ratio close to unity. Similarly, other planetary parameters, such as the surface pressure (P$_0$), surface temperature (T$_0$), and the radius of the planet (R$_p$) also begin to drift away from the true values, and become less constrained, at REC ratio $> 10^{-3}$. At REC ratio close to unity, the retrieval results suggest that the planet is $\sim50\%$ larger in size, with possibly only a tenuous atmosphere and smaller cloud fraction than the truth. Conversely, in the F$_\mathrm{ez}^\prime$/F$_p$ = $10^{-2}$ case, the added residual exozodi continuum is interpreted by the retrieval as a planet that is $\sim5\%$ smaller than the truth, with 50\% higher cloud coverage. These results and potential points of confusion highlight the critical need to account for exozodi either before the retrieval step by applying a removal algorithm, or perhaps including parameters describing the exozodi (e.g. number of zodis, exozodi color/composition, and possibly spatial information) within the retrieval itself. We leave these approaches open to future studies.

\subsubsection{Archean Earth atmosphere}
Our analysis of the Archean Earth atmosphere yields a similar result to the modern Earth atmosphere. The CH$_4$ and CO$_2$ bands, however, have band depths of $\sim10-15\%$ relative to the continuum including 1 zodi of gray-scattering exozodi. Shorter wavelength absorption bands will require PPF$_\mathrm{ez}$ values closer to $10^2$ for systems with a few zodis of dust, and $10^3$ for longer wavelength absorption bands, if molecular abundance measurement is the ultimate goal; this requirement may be relaxed by an order of magnitude or more if a binary result is sufficient. Higher spectral resolution observations may help increase the detectability of some spectral features (see Section~\ref{sec:specres_disc}). Plots of the Archean analysis are not shown here; the qualitative behavior parallels that of the modern Earth case.









\section{Discussion}\label{sec:mitigation}
Here we consider the exozodi problem in the context of our current knowledge of exozodi, post-processing capabilities, and HWO mission development. 

\subsection{Observational considerations}

For a given observation (e.g. spectral feature), minimizing the flux ratio F$_\mathrm{ez}^\prime$/F$_p$ through a combination of careful target choice and the application of an exozodi subtraction routine will maximize our ability to measure planetary and atmospheric properties. Obviously, nearby targets will maximize the denominator of the exozodi-to-planet flux ratio, especially if clouds are present in the atmosphere to reflect more starlight. Minimizing the numerator, however, will present a data processing challenge. This is especially true for systems with high dust densities and/or gray-to-red dust, where precise post-processing techniques will be required to remove exozodi to the $\lesssim 0.1\%$ level for precise atmospheric retrieval. Previous studies have made the reasonable assumption that other stellar systems have dust properties similar to the slightly reddened dust in our solar system \citep[e.g.][]{ertelHOSTSSurveyExozodiacal2020}. Given the results of this study, zodi-like dust may significantly impede our ability to characterize exoEarths if it is not precisely removed from the observations, which may be challenging for nearby Sun-like systems, which have $>3$ zodis of dust \citep[][]{ertelHOSTSSurveyExozodiacal2020}. Upcoming space based missions (e.g. the Nancy Grace Roman Space Telescope) and new ground-based instrumentation may better constrain exozodi density, spatial, and composition information in the time before HWO launches \citep[e.g.][]{currie2025exozodiacal}. 

\subsection{Exozodi subtraction capabilities}
Exozodi removal algorithms have been developed to subtract exozodiacal dust from directly imaged exoEarth systems to the Poisson noise limit, but are currently limited. For smoothly-varying exozodi, \citet{kammererSimulatingReflectedLight2022} show that it is possible to remove exozodi down to the photon-noise limit through parametric modeling for face-on systems with up to 1000 zodis of dust, or up to $\sim50$ zodis for systems inclined $60^\circ$ from face-on. Because the photon-noise limit is not a fixed fractional residual, but instead depends on the exozodi brightness and integration time, whether it satisfies the requirements quantified in this work will depend on the dust density, inclination, and structure of a given system \citep[][]{kammererSimulatingReflectedLight2022, currieMitigatingWorstcaseExozodiacal2023}. However, systems that are $> 60^\circ$ inclined from face-on, which represent $\sim50\%$ of targets if we assume random orientations, will pose a larger challenge. Fully removing spatially inhomogeneous exozodi may only be possible for systems with up to $\sim 3$ zodis and $<60^\circ$ inclination with a modern Gaussian convolution technique \citep[][]{currieMitigatingWorstcaseExozodiacal2023}. Subtracting exozodi to the $\lesssim0.1\%$ level regardless of inclination, density, or structure, especially at wavelengths longer than $> 1 \mu$m remains an open area of research.

\subsection{Impact on HWO mission development}

While exozodi has been a consideration for HWO yield studies \citep[][]{morgan2023exo, stark2024paths} and some precursor science \citep[e.g.][]{kammererSimulatingReflectedLight2022, currieMitigatingWorstcaseExozodiacal2023, mennesson2024current}, this work has quantified the exozodi problem in greater detail, providing novel insights that can impact the design, development, and execution of the HWO mission. Of course, increasing the primary mirror diameter is perhaps the most mathematically straightforward way to reduce the impact of exozodi since it allows us to ``resolve out'' extended background sources. Perhaps a more practical approach is to design the fiducial exoEarth survey to target systems with lower inclinations and lower exozodi densities; surveys with JWST, Roman, and ground-based facilities may reveal more information on exozodiacal dust populations in nearby Sun-like systems in the time before HWO launches \citep[e.g.][]{mennesson2019interplanetary, mennesson2019potential, douglasSensitivityRomanCoronagraph2022, currie2025exozodiacal}. 



\label{sec:specres_disc}
\begin{figure}
    \centering
    \includegraphics[width=\linewidth]{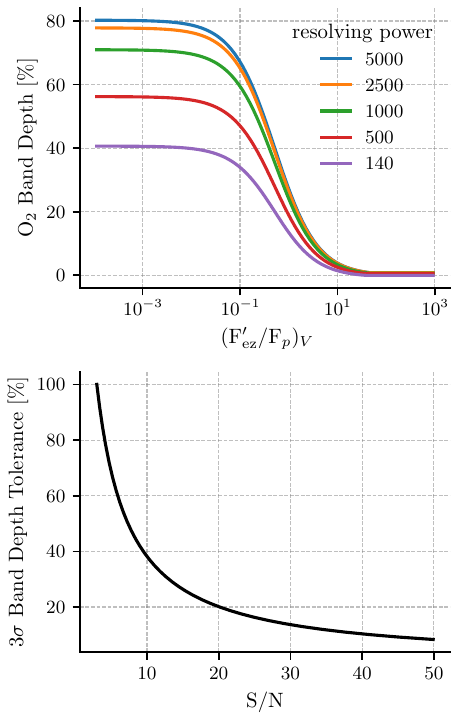}
    \caption{O$_2$ band depth as a function of $(F_\mathrm{ez}^\prime/F_\mathrm{p})_V$ and spectral resolution (upper), and the band depth required for a 3$\sigma$ detection as a function of spectral bin S/N (lower), assuming S/N is equal across all spectral bins. Increasing spectral resolution is one method to ease exozodi subtraction and S/N requirements for molecular detection. }
    \label{fig:bandtolerance}
\end{figure}

\begin{figure}
    \centering
    \includegraphics[width=1\linewidth]{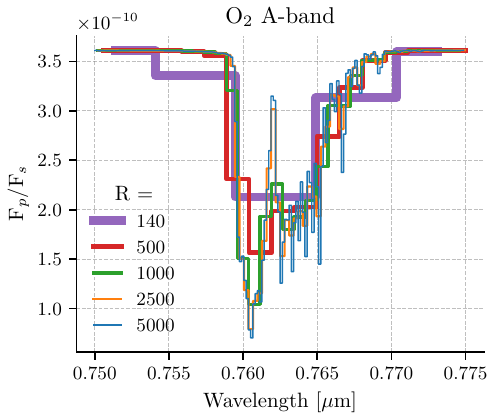}
    \caption{The O$_2$ A-band plotted at different spectral resolutions. Increasing spectral resolution also increases the effective absorption band depth relative to the continuum. }
    \label{fig:o2_resolutions}
\end{figure}

Increasing the spectral resolution of the instrument effectively captures finer band structure and deeper absorption lines, improving the apparent integrated depth of an absorption feature. We investigate this approach in Figure~\ref{fig:bandtolerance}, where we plot the O$_2$ band depth as a function of REC ratio for spectral resolutions between R=140--5000 in the upper panel. We then calculate the band depth required for a 3$\sigma$ detection as a function of S/N in the lower panel, optimistically assuming that the S/N of the continuum and within the band are equal. We plot the uncontaminated O$_2$ A-band spectra at these spectral resolutions in Figure~\ref{fig:o2_resolutions} for reference. 

Higher spectral resolution relaxes REC ratio requirements, and may also lower S/N requirements. For example, with an S/N=10 detection of the spectral bins, we require the absorption band depth to be $>40\%$ with respect to the continuum to detect the feature at 3$\sigma$ significance at R=140. This translates to subtracting the exozodi flux to within approximately 1\% of the planet's flux, necessitating a PPF$_\mathrm{ez} \gtrsim 1000$ for $N_\mathrm{ez} \gtrsim 10$ zodis of dust for a target 10 pc away (see Figure~\ref{fig:NezPPF}). However, increasing the spectral resolution by approximately an order of magnitude can relax this PPF$_\mathrm{ez}$ requirement by $\sim 30$x; since absorption bands at higher spectral resolution are deeper, achieving an observed 40\% band depth allows for a higher tolerance of exozodi contamination (Figure~\ref{fig:bandtolerance}). 

Consequently, achieving higher spectral resolution may require paying an exposure time penalty to maintain S/N = 10 in the spectral bins. However, this penalty can be offset by accepting lower S/N for the spectral bins since deeper bands may be easier to distinguish from the continuum. Continuing with the example in the previous paragraph, increasing the spectral resolution from R=140 to R=1000 means that the depth of the O$_2$ band is instead $60\%$ with respect to the continuum when the exozodi is subtracted to within 10\% of the planet's flux. In this case, only S/N$\sim5$ is required for the spectral bins (Figure~\ref{fig:bandtolerance}).
In summary, if a simple binary detection result is sufficient, increasing the resolving power of the instrument may be a promising approach to balance the detection threshold with S/N requirements. 

\section{Conclusions}\label{sec:conclusion}

Extracting exoEarth spectra from HWO observations may be complicated by exozodiacal dust. 
To minimize the exozodi spectral effect, our results suggest that the exozodi must be removed from extracted exoEarth spectra such that the residual exozodi-to-planet flux ratio in the V-band is $<10^{-2}$ to preserve band depths, and $<10^{-3}$ to retrieve molecular abundances. Because this requirement is defined relative to the planet, it is independent of the dust level; expressed instead as a fraction of the original exozodi flux, the required removal scales inversely with the exozodi brightness. For a system in which the exozodi is comparable to the planet in the V-band ($\sim10$ zodis at 10~pc), retrieving abundances requires removing all but $\sim0.1\%$ of the original exozodi flux, tightening in proportion to the dust density (e.g., $\sim0.01\%$ at $\sim100$ zodis). This will be increasingly challenging to achieve for high dust densities with modern techniques. However, if a binary detection is sufficient, this can lower the required exozodi post-processing factor by an order of magnitude or more, especially if spectral resolution can be increased.

While the exozodi spectral effect applies to any system with exozodiacal dust, some subtleties arise for different dust compositions. Exozodi that is predominantly red or gray in color, similar to our solar system's zodiacal dust, can introduce a strong bias in spectral interpretation. Characterizing exoEarths in systems with bluer dust, however, may be less limiting due to the weaker spectral effect at longer wavelengths. Color minimally impacts the shape of narrow molecular absorption bands, but has a stronger effect on the shape of broader features, like the Rayleigh slope. To characterize potentially dozens of Earth-sized exoplanets, a driving science goal for the HWO mission, it will be critical to also understand the exozodiacal dust in these systems. 


\begin{acknowledgments}
The authors thank the reviewer for their thoughtful comments and suggestions that strengthened the quality of this manuscript. Additionally, the authors thank John Debes for helpful discussions on debris disk colors. 
The work of M.H.C. and E.A. was supported by appointments to the NASA Postdoctoral Program at the NASA Goddard Space Flight Center, administered by Oak Ridge Associated Universities under contract with NASA (ORAU-80HQTR21CA005). 
\end{acknowledgments}

\appendix
\section{Retrieval Results}\label{apx:cornerplots}
\begin{figure*}
    \centering
    \includegraphics[width=\linewidth]{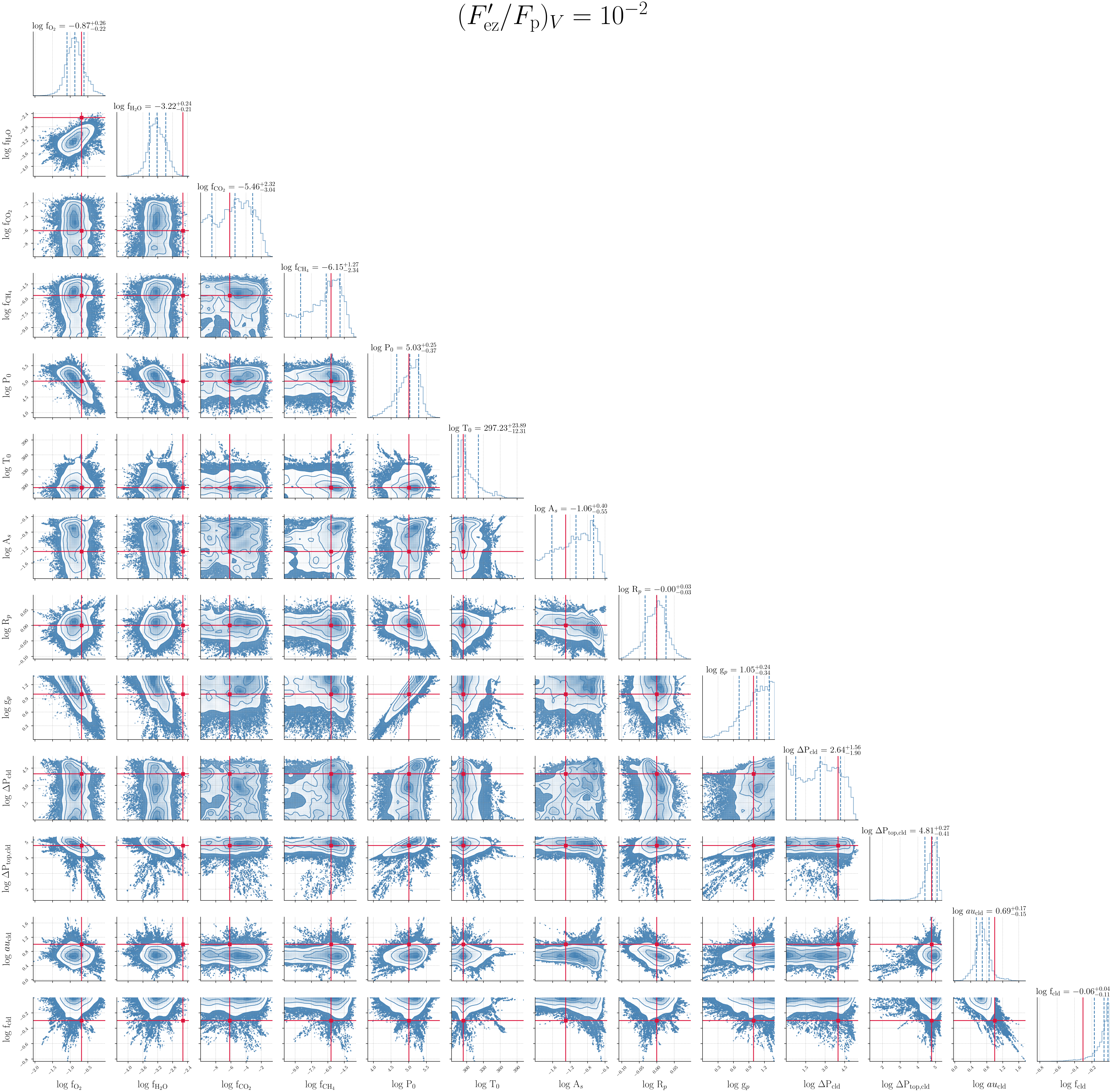}
    \caption{Corner plot for retrieving properties of a modern Earth twin with 50\% clouds at 10 pc away, with an exozodi to planet flux ratio  $(F_\mathrm{ez}^\prime/F_\mathrm{p})_V = 10^{-2}$. The corner plots associated with all five $(F_\mathrm{ez}^\prime/F_\mathrm{p})_V$ cases are available in the online journal.}
    \label{fig:corner}
\end{figure*}

Here we present corner plots associated with the retrievals from Section~\ref{sec:retrievalresults} as a figure set available in the online journal. We show an example corner plot for the $(F_\mathrm{ez}^\prime/F_\mathrm{p})_V = 10^{-2}$ case in Figure~\ref{fig:corner}.








\bibliography{exozodi_paper}
\bibliographystyle{aasjournal}

\end{document}